\documentclass[aps,twocolumn,pre,showpacs,superscriptaddress]{revtex4-1}
\usepackage{graphicx}
\usepackage{amsmath}
\usepackage{amsfonts}
\usepackage[table]{xcolor}
\usepackage[caption=false]{subfig}
\usepackage{array}
\usepackage{floatrow}
\usepackage{pbox}
\usepackage{physics}
\usepackage[colorlinks=true,linkcolor=blue,citecolor=blue,filecolor=cyan,urlcolor=blue,breaklinks=true]{hyperref}

\newcommand{\mrd}{\mathrm d}
\newcommand{\mre}{\mathrm e}
\newcommand{\mean}[1]{\left\langle #1 \right\rangle}
\newcommand{\Var}{\mathrm{Var}}
\newcommand{\pstat}{p^\mathrm{s}}
\newcommand{\sm}{s_\mathrm{m}}
\newcommand{\stot}{s_\mathrm{tot}}

\begin{document}
\title{Finite-time generalization of the thermodynamic uncertainty relation}

\author{Patrick Pietzonka}
\affiliation{ II. Institut f\"ur Theoretische Physik, Universit\"at Stuttgart,
  70550 Stuttgart, Germany}
\author{Felix Ritort}
\affiliation{Departament de F\'isica Fonamental, Universitat de Barcelona,
  Diagonal 647, 08028 Barcelona, Spain}
\affiliation{CIBER-BBN de Bioingenier\'ia, Biomateriales y Nanomedicina, Instituto de Salud Carlos III, 28029 Madrid, Spain}
\author{Udo Seifert}
\affiliation{ II. Institut f\"ur Theoretische Physik, Universit\"at Stuttgart,
  70550 Stuttgart, Germany}
\date{\today}

\parskip 1mm

\begin{abstract}

  For fluctuating currents in non-equilibrium steady states, the recently
  discovered thermodynamic uncertainty relation expresses a fundamental
  relation between their variance and the overall entropic cost associated with the
  driving. We show that this relation holds not only for the long-time limit
  of fluctuations, as described by large deviation theory, but also for
  fluctuations on arbitrary finite time scales. This generalization facilitates
  applying the thermodynamic uncertainty relation to single molecule
  experiments, for which infinite timescales are not accessible. Importantly,
  often this finite-time variant of the relation allows inferring a bound on the
  entropy production that is even stronger than the one obtained from the long-time
  limit. We illustrate the relation for the fluctuating work that is performed
  by a stochastically switching laser tweezer on a trapped colloidal particle.

\end{abstract}

% Explanation of PACS numbers:
% 05.70.Ln: Nonequilibrium and irreversible thermodynamics
% 05.40.-a: Fluctuation phenomena, random processes, noise, and Brownian motion

\maketitle

\section{Introduction}

The arguably most prominent characteristics of a thermal system driven into a
non-equilibrium steady state (NESS) is its rate of entropy production
$\sigma$, i.e., the amount of heat that is transferred to a heat bath
per unit of time.  For an exact experimental determination of $\sigma$,
however, one would have to measure either the temperature change of a large
but yet finite heat bath or to keep track of the net (free) energy input of
all the driving forces. For micro- and nano-systems, such as present in single
molecule or soft matter experiments \cite{rito06,cili10,seif12}, the temperature changes of a macroscopic
heat bath are by far too small for the first method to be feasible. The second
method is viable only if the system is driven by mechanical forces acting on
observable degrees of freedom or for small electronic circuits \cite{peko15,gasp15}. However, the
quantitative energetic input of chemical driving maintained by macroscopic
particle reservoirs, so-called chemostats, is not yet accessible on a
molecular scale.

A lower bound on $\sigma$ can be inferred from the recently discovered
thermodynamic uncertainty relation by measuring the mean and variance of an
arbitrary non-vanishing current in a NESS \cite{bara15,ging16}. Turning the
argument around, in situations where $\sigma$ \textit{is} directly accessible,
the thermodynamic uncertainty can be used to predict the minimal variance of
any current. So far, this relation has been understood in the context of large
deviation theory \cite{touc09,bara15d,piet15,ging16,ging16a}, which has led to
refinements \cite{piet16,pole16} and variants for the diffusion in periodic
potentials \cite{tsob16}, stochastic pumps \cite{rots16}, and first passage
problems \cite{garr17}. Moreover, the thermodynamic uncertainty relation has
been considered theoretically in such diverse contexts as enzyme kinetics
\cite{bara15a}, self-propelled particles \cite{fala16}, magnetic systems
\cite{guio16}, self-assembly \cite{nguy15}, Brownian clocks \cite{bara16}, and
the efficiency of molecular motors \cite{piet16b}.

The thermodynamic uncertainty relation as established so far, crucially
relying on large deviation theory, considers fluctuations that occur in the
limiting case of large time scales. Estimating large deviation functions
experimentally is possible on the basis of large sets of data and if the
probability of untypical fluctuations decays slowly enough to make the long
time limit accessible \cite{mart16}. In contrast, the theory of stochastic
thermodynamics \cite{seif12} has proven most fruitful for experimental
applications in cases where it provides relations that hold on \textit{finite}
time scales. Most prominently, the Jarzynski relation \cite{jarz97} and the
Crooks fluctuation theorem \cite{croo99} allow one to infer free energy
differences from the measurement of the fluctuating work during finite-time
protocols, see, e.g., \cite{coll05}. Similarly, the concept of
stochastic entropy \cite{seif05a} allows for a generalization of the detailed
fluctuation theorem for the entropy production in a NESS \cite{lebo99} to
finite and thus experimentally accessible time scales \cite{spec07}.

In this paper, we show, based on extensive numerical evidence, that the
thermodynamic uncertainty relation can be generalized to fluctuations on
finite time scales as well. We illustrate this finite-time version with
experimental data for fluctuations of work performed on a colloidal particle
in a dichotomously switching trap \cite{gome10,diet15a,wang16}. This
illustration serves as a proof of principle for applying the uncertainty
relation in the future to more complex experimental systems with more than one
input or output current such as Brownian heat engines \cite{blic12,mart16} and
molecular motors, see, e.g., \cite{viss99}.  For small electronic
circuits at low temperature this approach may become complementary to the
recent progress in calorimetrically measuring heat transfer
\cite{peko15,gasp15}.

The paper is organized as follows. In Sec.~\ref{sec:mainres} we state the main
result, which is then illustrated experimentally in Sec.~\ref{sec:exp}. In
Sec.~\ref{sec:genf} the result is put on a theoretical basis, conjecturing a
bound on the generating function for currents. This bound is illustrated and
verified numerically and proven for the limits of short times and linear
response. We conclude in Sec.~\ref{sec:conclusion}.

\section{Main result}
\label{sec:mainres}
For a thermodynamic system modeled as a Markovian network and driven into a
NESS by time independent forces, we consider the
fluctuations of an arbitrary time-integrated current $X(t)$ with
$X(0)=0$. While the average of such a current increases linearly in time $t$
as
\begin{equation}
  \mean{X(t)}\equiv Jt,
\end{equation}
where $\mean{\cdots}$ denotes the steady-state average, other characteristics
of the distribution of $X(t)$ typically exhibit a more complex dependence on the
observation time $t$. For the variance $\Var[X(t)]\equiv\mean{X(t)^2}-\mean{X(t)}^2$, we demonstrate that
\begin{equation}
  \Var[X(t)]\sigma/J^2t\geq  2k_\mathrm{B}
  \label{eq:mainres}
\end{equation}
holds for arbitrary times $t>0$, where $k_\mathrm{B}$ is Boltzmann's
constant. Thus, the fluctuations of $X(t)$ at finite times can be related to the
rate of total entropy production $\sigma$ associated with the driving. 
In the limit of large observation times the variance of $X(t)$ settles to a
linear increase with the effective diffusion coefficient
$D\equiv\lim_{t\to\infty}\Var[X(t)]/2t$.
On this infinite time scale, the uncertainty relation reads
$D\sigma/J^2\geq k_\mathrm{B}$, which has previously been reported
\cite{bara15} and proven \cite{ging16}.

\section{Experimental illustration}
\label{sec:exp}
\begin{figure}
  \centering
  \includegraphics[width=0.85\textwidth]{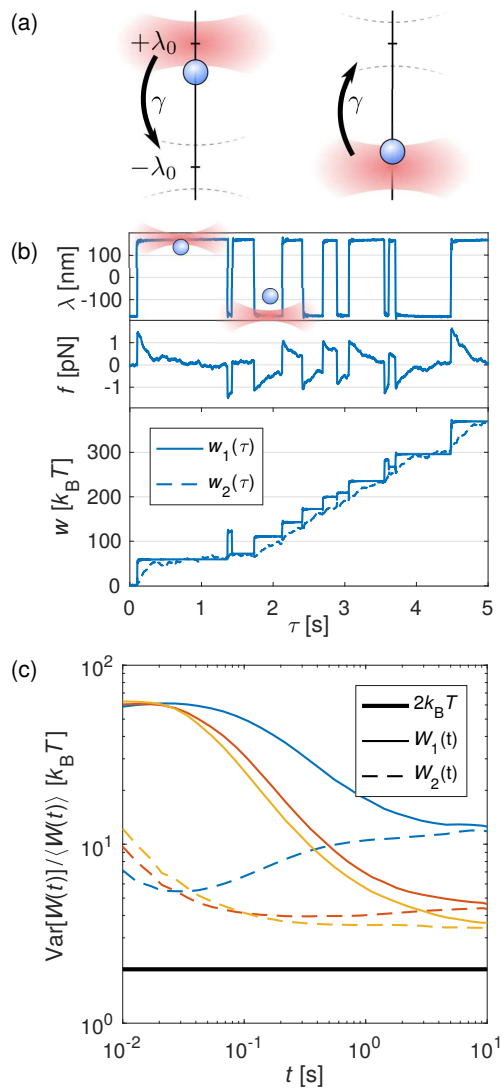}
  \caption{Experimental data for the bound~\eqref{eq:mainres} for a
    colloidal particle in a stochastically switching trap, as
    sketched in (a). Panel (b) shows the time dependent position $\lambda(\tau)$ of the trap, the force $f(\tau)$ exerted on the
    colloid and the work $w_{1,2}(\tau)$ according to the two definitions~\eqref{eq:workdef}
    for a short part of the trajectory. In (c), the quantity
    $\Var[w(t)]/\mean{w(t)}$ is shown as a function of the length $t$
    of the time interval and compared to the lower bound $2k_\mathrm{B}
    T$. Data refer to the amplitude $\lambda_0\simeq170\,\mathrm{nm}$ and a trap
    with inverse relaxation time $\tau_\mathrm{rel}^{-1}\simeq 4.6\,\mathrm{s}^{-1}$ throughout. For the blue lines the switching rate is
    $\gamma\simeq 2.88\,\mathrm{s}^{-1}$. In (c), we show additional
    data for $\gamma\simeq 8.73\,\mathrm{s}^{-1}$ (red) and
    $\gamma\simeq 12.3\,\mathrm{s}^{-1}$ (yellow).}
  \label{fig:experiment}
\end{figure}

As an experimental illustration of the relation~\eqref{eq:mainres}, we
analyze data for a colloidal particle in a dichotomously switching optical
trap \cite{diet15a}. The center of the trap is switched along a one-dimensional coordinate $\lambda(\tau)$ between the positions $+\lambda_0$ and
$-\lambda_0$ at points in time that are generated by a Poisson process with
rate $\gamma$ [see Fig.~\ref{fig:experiment}(a)]. The force $f(\tau)$ which is exerted on the bead along this
dimension is measured directly from the deflection of the light. We consider
two different definitions of work \cite{schu03,moss09a}
\begin{subequations}
\begin{equation}
  w_1(\tau)\equiv\int_0^\tau\mrd\lambda(\tau')\,f(\tau')\approx\sum_{n=1}^{\tau/\delta\tau}(\lambda_n-\lambda_{n-1})\frac{f_n+f_{n-1}}{2}
\end{equation}
and
\begin{equation}
  w_2(\tau)\equiv-\int_0^\tau\mrd f(\tau')\,\lambda(\tau')\approx-\sum_{n=1}^{\tau/\delta\tau}(f_n-f_{n-1})\frac{\lambda_n+\lambda_{n-1}}{2}.
\end{equation}
\label{eq:workdef}%
\end{subequations}
The discrete integration schemes with $f_n\equiv f(n\,\delta\tau)$ and
$\lambda_n\equiv \lambda(n\,\delta\tau)$ define the integrals for
discontinuous $\lambda(\tau)$ and $f(\tau)$ via the limit $\delta\tau\to 0$
and are used to compute the work for experimental data captured with a finite
time resolution of $\delta \tau\simeq 1\,\mathrm{ms}$.  We interpret
$w_1(\tau)$ as the work performed by moving the trap against the force
$f$. The second definition, $w_2(\tau),$ is equivalent to $w_1(\tau)$ up to a
finite boundary term of the form $\lambda f$.  Figure~\ref{fig:experiment}b
shows sample data for $\lambda(\tau)$ and $f(\tau)$ together with
$w_{1,2}(\tau)$.

Due to the stochastic switching of the trap, the system reaches a NESS for
long observation times $\mathcal{T}$. Hence, the steady state averages and
cumulants for the work $W_{1,2}(t)\equiv w_{1,2}(\tau)-w_{1,2}(\tau-t)$
performed on finite time intervals $t\ll\mathcal{T}$ can be obtained from the
time average over $\tau\in[t,\mathcal{T}]$.

\begin{figure*}
  \centering
  \includegraphics[width=\linewidth]{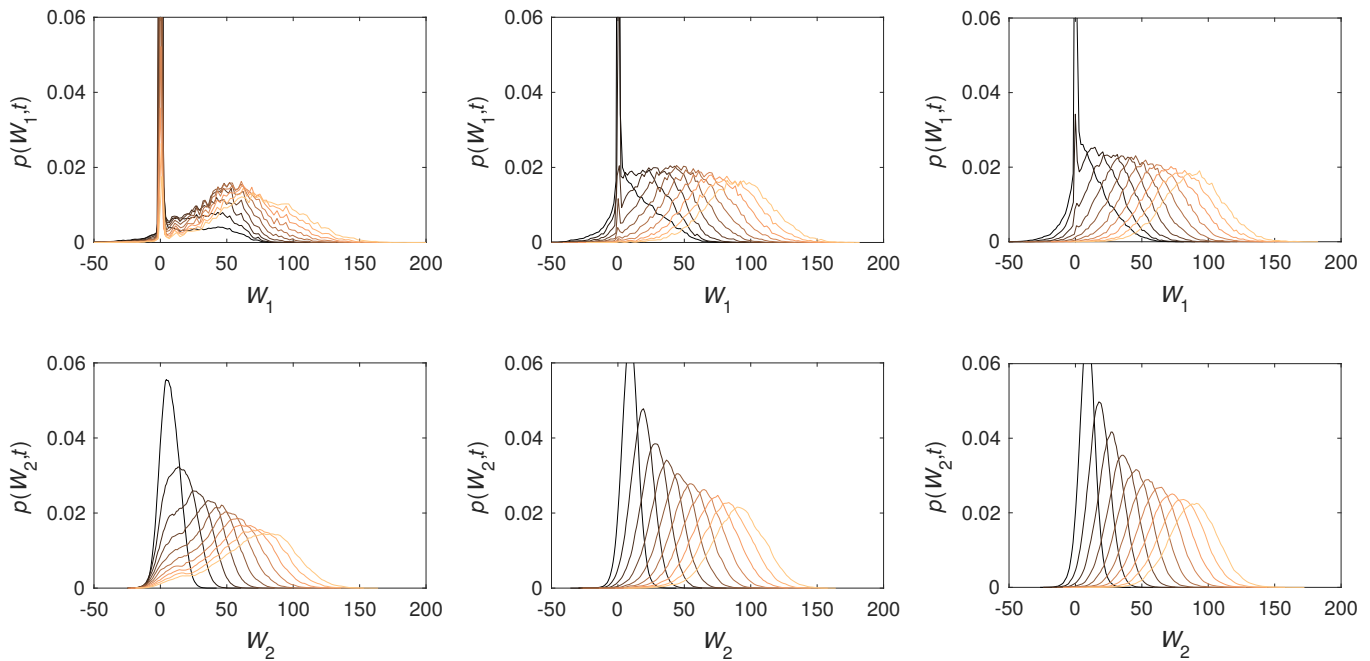}
  \caption{Full distributions of the work underlying the data for mean and
    variance in Fig.~\ref{fig:experiment}(c). Left column: $\gamma\simeq
    2.88\,\mathrm{s}^{-1}$ [blue in Fig.~\ref{fig:experiment}(c)], middle column $\gamma\simeq
    8.73\,\mathrm{s}^{-1}$ [red in Fig.~\ref{fig:experiment}(c)], and right column $\gamma\simeq
    12.3\,\mathrm{s}^{-1}$ [yellow  Fig.~in \ref{fig:experiment}(c)]. Time $t$ increases from
    $0.1\,\mathrm{s}$ (black) to $1\,\mathrm{s}$ (light brown) in steps of $0.1\,\mathrm{s}$.}
  \label{fig:workdist}
\end{figure*}
In Fig.~\ref{fig:workdist}, we show the full distributions of work performed
on the colloidal particle. Unlike for deterministic switching, these distributions can
be highly non-Gaussian at finite times. The time scale chosen in these plots
covers the transition from work fluctuations in a typically resting trap for
short times to work fluctuations that are directly affected by switching the
trap.  Since the work $W_1$ increases in a step-like fashion [see
Fig.~\ref{fig:experiment}(b)], its distribution exhibits a sharp peak
corresponding to time intervals where the trap does not switch. With
increasing length of the time interval the height of this peak decreases and a
second bulge in the distribution starts growing. This part of the distribution
is much broader since the work performed while switching the trap is
stochastic. For the work $W_2$, fluctuations occur also while the trap is at
rest, leading to a broader peak at short times. With increasing switching rate
$\gamma$ of the trap, the effects of the resting trap become less pronounced,
leading to an overall smoother work distribution. 

For long time intervals $t$, both definitions of the work measure the area
enclosed by the trajectory in the $(\lambda,f)$ space up to a finite
contribution that does not scale with $t$. Thus, in the long-time limit,
cumulants of $W_1(t)$ become equal to the respective cumulants of $W_2(t)$ to
leading order in time. In particular, as the mean is independent of $t$, we
have $\mean{W_1(t)}/t=\mean{W_2(t)}/t=\sigma T$, where $T$ is the temperature
of the surrounding heat bath.  Since the work that is performed on the system
must ultimately be dissipated, we can indeed identify these averages with the
rate of entropy production $\sigma$. Thus, specifying $W_{1,2}(t)$ as
integrated current in~\eqref{eq:mainres}, we obtain the bound
\begin{equation}
  \frac{\Var[W_{1,2}(t)]}{\mean{W_{1,2}(t)}}\geq 2k_\mathrm{B}T
  \label{eq:ur_work}
\end{equation}
on the fluctuations of $W_{1,2}$. As Fig.~\ref{fig:experiment}c shows, this
bound is satisfied for arbitrary times $t$, various values of the switching
rate $\gamma$, and for both definitions $W_1(t)$ and $W_2(t)$. In the limit of
large $t$, for which the uncertainty relation has previously been shown to
hold, the expression on the left-hand side of Eq.~\eqref{eq:ur_work} becomes
equal for both definitions. In contrast, for finite time intervals the
fluctuations of $W_1(t)$ and $W_2(t)$ differ by a whole order of
magnitude. Thus, the finite-time generalization of the uncertainty relation
allows one to infer stronger lower bounds on the entropy production by choosing
the most suitable among various currents that become equivalent in the
long-time limit. Most remarkably, the difference to the bound can be smaller
for finite times than it is in the long-time limit, as the minimum of the blue
dashed curve in Fig.~\ref{fig:experiment}, corresponding to a slow switching
rate $\gamma$, shows.  The finite-time bound evaluated at $t\simeq
0.03\,\mathrm{s}$ yields $5.4\,k_\mathrm{B}T$ and is thus about a factor of
$2$ better than the long-time value $12.0\,k_\mathrm{B}T$.

The relation between the variance and mean of work fluctuations has previously
been discussed for transient non-equilibrium processes \cite{rito04,funo16}.
For those, it is possible to obtain a ratio of these quantities that is
smaller than the bound set by Eq.~\eqref{eq:ur_work}, which applies to steady
states.

\section{Bound on the generating function}
\label{sec:genf}
\subsection{General formulation}
In the following, we discuss the evidence for the finite-time
bound~\eqref{eq:mainres} in a broader theoretical framework.
We represent the system as a set of states $\{i\}$ and Markovian transition
rates $k_{ij}\geq 0$ from state $i$ to state $j$ and denote the corresponding
stationary distribution as $\pstat_i$.
A time-integrated current $X(t)$ is defined by specifying its change $d_{ij}=-d_{ji}$ upon a
transition from $i$ to $j$. The steady-state average of this current is
\begin{equation}
  J=\mean{\dot X(t)}=\sum_{ij}\pstat_i k_{ij}d_{ij}.
  \label{eq:ssavg}
\end{equation}
In particular, the choices
\begin{equation}
  d_{ij}^\mathrm{m}\equiv\ln\frac{k_{ij}}{k_{ji}}
\qquad \mathrm{and}\qquad 
  d_{ij}^\mathrm{s}\equiv\ln\frac{\pstat_ik_{ij}}{\pstat_jk_{ji}}
\label{eq:ds}
\end{equation}
define the entropy production in the medium $\sm(t)$ and the total entropy
production $\stot(t)$, respectively, which are rendered dimensionless by
setting $k_\mathrm{B}=1$ here and in the following \cite{seif05a}. The steady state averages
\eqref{eq:ssavg} of these two currents are equal, defining the entropy
production rate
$\sigma\equiv \mean{\dot s_\mathrm{m}}=\mean{\dot s_\mathrm{tot}}.$
The fluctuations of any current $X(t)$ can conveniently be analyzed in terms of
the generating function
\begin{equation}
  g(z,t)\equiv\mean{\mre^{zX(t)}}=\mel**{1}{\mre^{t\mathcal{L}(z)}}{\pstat}
  \label{eq:gdef}
\end{equation}
with the tilted transition matrix
\begin{equation}
  \mathcal{L}_{ij}(z)\equiv k_{ji}\exp(zd_{ji})-\delta_{ij}\sum_\ell k_{i\ell}
  \label{eq:tilted}
\end{equation}
and the vector $\bra{1}$ containing $1$ in every entry.
This function allows one to infer the mean of the current as
\begin{equation}
  \mean{X(t)}=\left.\partial_z\ln g(z,t)\right|_{z=0}
\end{equation}
and its variance as
\begin{equation}
  \Var[X(t)]=\left.\partial_z^2 \ln g(z,t)\right|_{z=0}.
  \label{eq:varg}
\end{equation}

In extensive numerical checks described below, we find that the logarithm of
the generating function satisfies the parabolic lower bound
\begin{equation}
(1/t) \ln g(z,t)\geq J z(1+z J/\sigma),
  \label{eq:gbound}
\end{equation}
which is our most general theoretical result.  In the limit $t\to\infty$, the
left-hand side of this expression converges to the Legendre transform of the
large deviation function associated with the current $X(t)$. In this limit,
the parabolic bound has been conjectured in \cite{piet15} and proven in
\cite{ging16}. Our new finding generalizes this result to the regime of
fluctuations on finite time scales, which are inherently not accessible by
large deviation theory. Crucially, the difference between $(1/t)\ln g(z,t)$ and
the parabolic bound can be smaller for finite times $t$ than it is in the
long-time limit. Such a behavior of the generating function is necessary for a
minimum of the ratio $\Var[X(t)]/\mean{X(t)}$ at finite time $t$ as in our
experimental illustration in Fig.~\ref{fig:experiment} for the work $W_2(t)$
at low switching rate.

The bound~\eqref{eq:gbound} is globally saturated for a Gaussian distribution
of the current \footnote{Even though the distribution of work $W_2$ often looks
Gaussian in our experimental case study (see Fig.~\ref{fig:workdist}), the
generating function would in these cases not be parabolic due to non-Gaussian
tails of the distribution. Accordingly, the uncertainty relation is not saturated.}, as
observed for a biased diffusion in a flat potential. This process can be
approximated by a discrete asymmetric random walk on a ring where the number
of states is let to infinity while the affinity per step is let to
zero. Otherwise, the bound is only trivially saturated for $z=0$ and, as a
consequence of the fluctuation theorem \cite{seif05a}, for the generating
function of $\stot$ at $z=-1$. For other currents that become equal to $\stot$
on large time scales, such as the medium entropy production $\sm$, the bound
is approached at $z=-1$ only in the long-time limit.

Of experimental relevance is mainly the variance~\eqref{eq:varg} of the
current $X(t)$. Since $(1/t) \ln g(z,t)$ touches the bound at $z=0$ for
all $t$, the finite-time version~\eqref{eq:mainres} of the thermodynamic
uncertainty relation follows from the relation~\eqref{eq:gbound}.

\subsection{Illustration for unicyclic networks}

As a simple example, for which the generating function can be calculated explicitly,
we consider the asymmetric random walk on a ring with $N$ states and uniform
forward and backward transition rates $k^+$ and $k^-$. For the current averaged along all
links, the tilted transition matrix~\eqref{eq:tilted} reads 
\begin{equation}
  \mathcal{L}_{ij}(z)=k^+\mre^{z/N}\delta_{i,j+1}+k^-\mre^{-z/N}\delta_{i+1,j}-(k^++k^-)\delta_{i,j},
\end{equation}
where we identify the states $N+1\equiv 1$. The average current is
$J=(k^+-k^-)/N$ and the entropy production is $\sigma=(k^+-k^-)\ln(k^+/k^-)$. The stationary distribution
$\pstat_i=1/N$ is an eigenvector of $\mathcal{L}(z)$ for every $z$, hence the
generating function~\eqref{eq:gdef} becomes
\begin{equation}
  g(z,t)=\exp\left[t\left(k^+\mre^{z/N}+k^-\mre^{-z/N}-k^+-k^-\right)\right].
\end{equation}
It can be easily checked that this generating function satisfies the bound
\eqref{eq:gbound} at all times $t$. The bound is saturated for small $z$ in the linear
response limit of vanishing affinity $\ln(k^+/k^-)$ per step.

\begin{figure}
  \centering
  \includegraphics[width=0.59\linewidth]{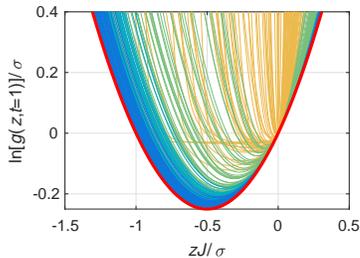}
  \caption{Generating function of the average current at time $t=1$ in a unicyclic network
    with perturbations of strength $\varepsilon\in\{0.05,0.1,0.5,5\}$ (from
    blue to orange). The unperturbed network has five states and rates
    $k^+=\mre^1$ and $k^-=1$. The bound~\eqref{eq:gbound} is shown as a red curve.}
  \label{fig:arw}
\end{figure}

We observe numerically that these unicyclic asymmetric random walks are
``optimal'' in the sense that they minimize the generating function at any given
$z$ and $t$. Changing the rates non-uniformly and adding further cycles only
increases the distance from the bound. In order to illustrate this observation, we show in
Fig.~\ref{fig:arw} the effects of perturbations of the rate matrix of the type
\begin{equation}
  k_{ij}=k^+\mre^{\varepsilon\theta_i^+}\delta_{i,j+1}+k^-\mre^{\varepsilon\theta_i^-}\delta_{i+1,j}+\varepsilon\phi_{ij},
\end{equation}
where the $\theta_i$ are independently drawn from a standard normal
distribution and the $\phi_{ij}$ are zero for $|i-j|\leq 1$ and
exponentially distributed otherwise. While the terms with $\theta_i^\pm$ make
the unicyclic rates non-uniform, the terms $\phi_{ij}$ add further cycles to
the network. We calculate the generating function numerically for $t=1$, which
qualifies as an intermediate time scale for transition rates of order $1$. The
bound~\eqref{eq:gbound} is satisfied in all cases.

\subsection{Short-time and linear response limits} While a full proof of the
parabolic bound~\eqref{eq:gbound} seems to be currently out of reach, we can
prove a weaker bound, which becomes equivalent to~\eqref{eq:gbound} for small
$t$. We start with the fluctuation relation
\begin{equation}
  p(-\stot,-X,t)/p(\stot,X,t)=\exp(-\stot)
\end{equation}
for the joint probability distribution of the total entropy production and the
current of interest at arbitrary time $t$, which follows directly from the time reversal of
the trajectories contributing to a fixed value of $\stot$ \cite{seif12}. Using this
relation, the generating function~\eqref{eq:gdef} can be written (dropping the
index `tot') as
\begin{align}
  g(z,t)&=\int\mrd s\int\mrd X\,p(s,X,t)\,\mre^{zX}=\frac{1}{2}\mean{\mre^{zX}+\mre^{-zX-s}}\nonumber\\
  &=\mean{\mre^{-s/2}\cosh(zX+s/2)}.
\end{align}
Bounding the hyperbolic cosine by a parabola that touches it at $z=0$ and $z=-s/X$, we obtain
\begin{align}
  g(z,t)\geq& 1+\mean{(1-\mre^{-s})zX(1+zX/s)}/2\nonumber\\
&=1+zJt+z^2\sigma t\int_0^\infty\mrd s\int_{-\infty}^{\infty}\mrd X
\psi(s,X)\,(X/s)^2\nonumber\\
&\geq 1+tJz(1+zJ/\sigma).
  \label{eq:linbound}
\end{align}
In the last step we have used Jensen's inequality for the averages with the
distribution \mbox{$\psi(s,X)\equiv p(s,X,t)s(1-\mre^{-s})/\sigma t$} for $s\geq 0$, which is
non-negative, normalized, and gives
\begin{equation}
  \int_0^\infty\mrd s\int_{-\infty}^{\infty}\mrd X\psi(s,X)\,(X/s)=J/\sigma.
\end{equation}

While the bound~\eqref{eq:linbound} is rigorous for arbitrary times $t$, it is useful mainly for
short times as a first order expansion of the otherwise stronger bound on
$g(z,t)$ that follows from Eq.~\eqref{eq:gbound}.
Indeed, for the variance of the current, the bound~\eqref{eq:linbound} implies for
arbitrary $t$
\begin{equation}
  \mean{X(t)^2}\geq  2tJ^2/\sigma .
  \label{eq:linres}
\end{equation}
Equation~\eqref{eq:mainres} differs from this relation only by the term
$\mean{X(t)}^2=J^2t^2$ and is thus proven for small times $t$ in linear order.

In the linear response regime for small driving affinity $\mathcal{A}$, the
current scales as $J\simeq \mathcal{A}$ and the entropy production rate as
$\sigma\simeq\mathcal{A}^2$. Hence the bound~\eqref{eq:linres} implies Eq.~\eqref{eq:mainres} in the linear
response limit for any fixed time, as follows from the scaling $J\sim
\mathcal{A}$ and $\sigma\sim\mathcal{A}^2$ for small driving affinities
$\mathcal{A}$.

\subsection{Numerical check for intermediate times}

\begin{figure}
  \centering
  \includegraphics[width=0.9\linewidth]{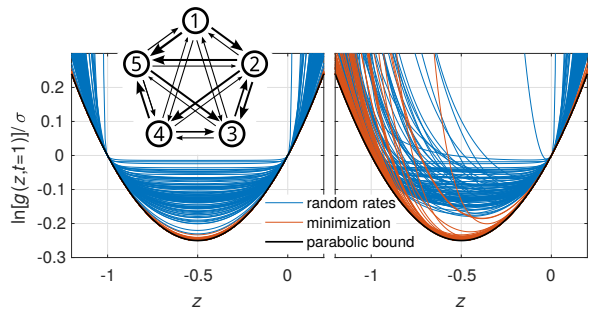}
  \caption{Numerical illustration of the bound on the generating function for
    a fully connected network with five states and random transition rates, as
    shown in the inset of the left panel and indicated by the different
    thicknesses of the arrows. We show generating functions $g(z,t=1)$
    for $\stot$ (left) and $\sm$ (right) calculated numerically for uniformly
    distributed $\ln k_{ij}\in[-5,5]$ and scaled by the entropy
    production rate $\sigma$ (blue). For each set of rates a local minimization of
    $g(z=-0.5,\,t=1)$ was performed, the corresponding generating functions are
    shown in red. In all cases, the bound $\sigma z(z+1)$ (shown in
    black) is satisfied.}
  \label{fig:numerics}
\end{figure}

On intermediate time scales we have verified the bound~\eqref{eq:gbound}
numerically using a combination of random search and optimization
techniques. At first, we have generated in total more than $3\times 10^5$
fully connected networks with $N\in\{3,4,5,7,10\}$ states and random
transition rates with $\ln k_{ij}$ distributed uniformly between $-12$ and
$5$. A sample of such a network with $N=5$ states is illustrated in the inset
of Fig.~\ref{fig:numerics}.  For these networks we have calculated the
stationary distribution and the generating function $g(z,t=1)$ via
Eq.~\eqref{eq:gdef} with $Jz/\sigma$ ranging from $-2$ to $1$. It is
sufficient to check the bound for $t=1$, since the large range of the choice
of transition rates effectively covers different timescales. This procedure
has been repeated for the currents of total entropy production
($d_{ij}=d_{ij}^\mathrm{s}$ in Eq.~\eqref{eq:ds}) and medium entropy
production ($d_{ij}=d_{ij}^\mathrm{m}$), as shown in Fig.~\ref{fig:numerics},
the current along an individual link $i\to j$, and a current defined by a
random asymmetric matrix $d_{ij}$. Each of the random networks has then been
used as a starting point for a constrained local minimization procedure that
varies the rates $k_{ij}$ to minimize $g(z,t=1)$ while keeping $\sigma$ and
$Jz/\sigma$ fixed (without this constraint the algorithm quickly finds the
linear response regime, for which we have proven the validity of the
bound). As Fig.~\ref{fig:numerics} illustrates for a small set of networks,
the bound~\eqref{eq:gbound} has proven valid for all of the random networks as
well as for the optimized networks.

\section{Conclusion} 
\label{sec:conclusion}
We have shown that the thermodynamic uncertainty relation between the
fluctuations of any current and the rate of entropy production in a NESS holds
on arbitrary timescales. This result follows from a parabolic bound on the
cumulant generating function associated with such a current. The fluctuation theorem
for entropy production allows proving this bound in the limit of short
timescales, complementing the previously known proof based on large deviation
theory for the long-time limit. For intermediate timescales the
bound is a conjecture that we have verified using extensive numerical
checks. A full proof in this regime seems to call for new mathematical
methods for the description of non-equilibrium steady states, which go beyond
fluctuation theorems and large deviation theory.

For an experimental illustration in the case where the entropy production is
measurable, we have analyzed this finite-time uncertainty relation with the
work that is performed on a colloidal particle in a stochastically switching
trap. As a next experimental step, it will be interesting to apply this
relation to systems driven by chemical reactions like molecular motors, in
order to bound the then \textit{a priori} unknown rate of entropy production
from below.  Our generalization of the thermodynamic uncertainty relation
should then become a valuable tool for inferring hidden thermodynamic
properties of driven systems from experimental trajectories of finite length.

\textit{Note added:} Recently, a proof of Eq.~\eqref{eq:mainres} for the
special case $X=\stot$ and Langevin dynamics has been reported in a preprint
\cite{pigo17}.\\ ~ \\

\begin{acknowledgments}
  We thank Andre C. Barato for fruitful interactions on
  related projects.
\end{acknowledgments}

\bibliography{/home/public/papers-softbio/bibtex/refs}

\end{document}